\documentclass[12pt,preprint,pra,aps]{revtex4}      
\usepackage{amsmath}  

\begin{document}  
\begin{flushright}
 \textbf{CERN-TH/2002-096} \\
\end{flushright}

\title{\vspace*{24pt} 
\LARGE\textbf{Quantum Memory:\\ Write, Read and Reset}\\[24pt]} 

\author{\large\textbf{Tai Tsun Wu}\vadjust{\kern10pt}} 

\affiliation{Gordon McKay Laboratory, Harvard University, 
Cambridge, Massachusetts, U.S.A.,}

\affiliation{Theoretical Physics Division, CERN, Geneva,
Switzerland\vadjust{\kern4pt}}

\author{\large\textbf{and}\\[16pt] Ming Lun Yu\vadjust{\kern10pt}}

\affiliation{41019 Pajaro Drive, Fremont, California,
U.S.A.\vadjust{\kern30pt}}
\altaffiliation{Present address: Applied Materials, Inc., 3050 
Bowers Ave., Santa Clara, California, U.S.A.
\vfil
\begin{flushleft}
 \normalsize \textbf{CERN-TH/2002-096} \\
 \textbf{June 2002}
\end{flushleft}
}   

\begin{abstract}
\begin{quote}
\hskip1.5em A model is presented for the quantum memory, the
content of which is a pure quantum state.  In this model, the fundamental
operations of writing on, reading, and resetting the memory are performed
through scattering from the memory.  The requirement that the quantum memory
must remain in a pure state after scattering implies that the scattering is of
a special type, and only certain incident waves are admissible.  An example,
based on the Fermi pseudo-potential in one dimension, is used to demonstrate
that the requirements on the scattering process are consistent and can be
satisfied.  This model is compared with the commonly used model for the
quantum memory; the most important difference is that the spatial dimensions
and interference play a central role in the present model.
\end{quote}
\end{abstract} 
\maketitle 
\thispagestyle{empty}
\eject

(1) It is the purpose of this paper to propose a model for quantum memory
(sometimes called a quantum register), which is an essential component of
any quantum computer.  While a classical memory contains a number, a quantum
memory contains a superposition of quantum states.  In this model, writing
on the quantum memory, reading the content of the quantum memory, and
resetting the quantum memory are all performed by scattering from the
quantum memory.  Because of this reliance on scattering
\cite{pikesabatier2002}, spatial dimension plays a central role.

Writing: The issue here is how a given superposition of quantum states
can be written on a quantum memory after the memory has been reset to a
standard state.

Reading: In this case the issue is how to extract from the memory whatever
superposition of states has been written there.

Resetting: For a quantum memory in an arbitrary superposition of
quantum states, it is desired to erase this information and replace it by a
standard state.  This should be done without violating unitarity and
time-reversal invariance.  It may be noted that while time reversal is not
an invariance of nature, its violation involves the second and third
generations of quarks \cite{christenson1964,Quark2,aubert&abe2001},
which are not expected to play any significant role in quantum memory.

(2) Let $|j\rangle$ be linearly independent quantum states for the memory; a
state for the memory is of the form $\sum_j a_j|j\rangle$.  Note that this
is a pure state, as the content of a quantum memory should be.  Without loss
of generality, let the standard state be $|1\rangle$.

In order to perform any operation on the memory, a wave or equivalently a
particle, or perhaps several particles, is sent in from the outside to
interact with the memory.  Thus the in field for the scattering process
is
\begin{equation}
\Psi^{\text{in}}=\left(\sum_j a_j^{\text{in}} |j\rangle\right)\psi^{\text{in}}.\label{eq:1}
\end{equation}
It should be emphasized that $\psi^{\text{in}}$ is our choice to accomplish
whatever the purpose of the scattering is.

The corresponding out state is in general a sum of products
\begin{equation}
\Psi^{\text{out}}=\sum_j|j\rangle\psi_j^{\text{out}}.\label{eq:2}
\end{equation}
This $\Psi^{\text{out}}$ contains information about the state of the memory
after scattering and the behavior of the scattered wave.  Later, the
scattered wave as described by $\psi_j^{\text{out}}$ moves away from the memory
and the information contained there is no longer available.  This means that
the final state of the memory is given by $\Psi^{\text{out}}$ \textit{averaged}
over this information that is no longer available.  On the other hand, in
order for the quantum memory to function, the final state of the memory
must be a pure state.  This condition of being a pure state leads to the
conclusion that $\Psi^{\text{out}}$ must also be of the form of Eq.\
(\ref{eq:1}), namely, Eq.\ (\ref{eq:2}) reduces to
\begin{equation}
\Psi^{\text{out}} =\left(\sum_j a_j^{\text{out}}|j\rangle\right)\psi^{\text{out}}.\label{eq:3} 
\end{equation} 
In other words, not only the in state but also the out state are
unentangled.  This leads to the important concept of admissible $\psi^{\text{in}}$.

A $\psi^{\text{in}}$ is defined to be \textit{admissible} \cite{footnote} if, for
all $\sum_j a_j^{\text{in}}|j\rangle$, the corresponding $\Psi^{\text{out}}$ is of
the form of Eq.\ (\ref{eq:3}).

(3) In order to perform the required operations, it is necessary to have a
sufficiently large collection of admissible $\psi^{\text{in}}$.  This
collection is sufficient if, given any $\sum_j a_j^{\text{in}}|j\rangle$ and
$\sum_j a_j^{\text{out}}|j\rangle$, there is at least one
\begin{equation}
\psi^{\text{in}}\left(\sum_j a_j^{\text{in}}|j\rangle\to \sum_j a_j^{\text{out}}|j\rangle\right),\label{eq:4}
\end{equation}
which has the property that, if Eq.\ (\ref{eq:1}) holds for this $\psi^{\text{in}}$, then Eq.\ (\ref{eq:3}) holds.

With this setup, the three fundamental operations of writing on, reading, and
resetting a quantum memory can be accomplished as follows.

(4) \textit{Writing}.  The simplest of these three operations is writing. 
Since the quantum memory has been reset beforehand, let $\sum_j
a_j^{(0)}|j\rangle$ be the given quantum state to be written on the memory. 
Then, writing can be accomplished by using $\psi^{\text{in}}\bigl(|1\rangle \to
\sum_j a_j^{(0)}|j\rangle\bigr)$.

(5) \textit{Reading}.  Reading from a quantum memory is very different from,
and more complicated than, writing on such a memory.  Consider a quantum
memory in a state $\sum_j a_j^{\text{in}}|j\rangle$.  It is desired to
determine the values of these coefficients $a_j^{\text{in}}$ by interrogating
this memory.  What this means is that we choose a sequence of admissible
$\psi^{\text{in}}$'s, i.e.,
\begin{equation}
\psi^{\text{in}(1)},\ \psi^{\text{in}(2)},\ \psi^{\text{in}(3)},\ \ldots,\ 
\psi^{\text{in}(N)},\label{eq:5}
\end{equation}
and then scatter them successively by this quantum memory.  More precisely,
consider the successive scattering processes:
\begin{equation}
\begin{array}{lclclccl}
&\sum_j a_j^{\text{in}}|j\rangle \psi^{\text{in}(1)}&\to& \sum_j
a_j^{\text{out}(1)}|j\rangle\psi^{\text{out}(1)}&\to& a_j^{\text{out}(1)}&=&a_j^{\text{in}(2)}\\[6pt]
\to &\sum_j a_j^{\text{in}(2)}|j\rangle\psi^{\text{in}(2)}&\to& \sum_j
a_j^{\text{out}(2)}|j\rangle\psi^{\text{out}(2)}&\to& a_j^{\text{out}(2)}&=&a_j^{\text{in}(3)}\\[6pt]
\to &\sum_j a_j^{\text{in}(3)}|j\rangle\psi^{\text{in}(3)}&\to& \sum_j
a_j^{\text{out}(3)}|j\rangle\psi^{\text{out}(3)}&\to& a_j
^{\text{out}(3)}&=&a_j^{\text{in}(4)}\\[4pt]  
\to&\cdots\\[4pt]
\to &\sum_j a_j^{\text{in}(N)}|j\rangle\psi^{\text{in}(N)}&\to&
\sum_j a_j^{\text{out}(N)}|j\rangle\psi^{\text{out}(N)}.\end{array}\label{eq:6}
\end{equation}
Corresponding to the list (\ref{eq:5}), we have the list of $\psi^{\text{out}}$'s, i.e.,
\begin{equation}
\psi^{\text{out}(1)},\ \psi^{\text{out}(2)},\ \psi^{\text{out}(3)},\ \ldots,\
\psi^{\text{out}(N)}.\label{eq:7}
\end{equation}
From the quantities listed in (\ref{eq:5}) and (\ref{eq:7}) together with
their interference, the values of $a_j^{\text{in}}$ are obtained.  Once the
$a_j^{\text{in}}$ are found, then the $a_j^{\text{out}(N)}$ can be calculated.  An
additional scattering using any one of the admissible $\psi^{\text{in}}\bigl(\sum_ja_j^{\text{out}(N)}|j\rangle\to\sum_j a_j^{\text{in}}|j\rangle\bigr)$ returns the quantum memory to its initial state.
 
(6) \textit{Resetting}.  With reading as prescribed above, resetting is now
straightforward.  Suppose we want to reset a quantum memory in the initial
state $\sum_j a_j^{\text{in}}|j\rangle$ to the standard state $|1\rangle$.  This
resetting consists of the following two steps.  (a) Read the memory as
described above.  [Note that, after this process of reading is performed, the
memory is in the original initial state $\sum_j a_j^{\text{in}}|j\rangle.$]  (b)
Apply an additional scattering using $\psi^{\text{in}}(\sum_j a_j^{\text{in}}|j\rangle\to |1\rangle)$.  This leaves the memory in the desired standard
state $|1\rangle$. 

(7) The concept of quantum memory introduced and described here is quite
general.  For example, the scattering process
\begin{equation}
\Psi^{\text{in}}=\left(\sum_j a_j^{\text{in}}|j\rangle\right)\psi^{\text{in}}
\to\Psi^{\text{out}}=\left(\sum_j a_j^{\text{out}} |j\rangle\right)\psi^{\text{out}}\label{eq:8}
\end{equation}
does not have many restrictions, and may or may not be linear.  Also the
linearly independent states $|j\rangle$ are allowed to depend on time, and may
or may not be the eigenstates of an operator.

Because of this generality, it is useful, and indeed necessary, to have some
simple examples.  One of the simplest such examples will be described below. 
Once this example is understood, it is not difficult to construct many other
examples.

(8) The example of the quantum memory to be described below relies on the
Fermi pseudo-potential in one dimension.  The original idea of the Fermi
pseudo-potential is to generalize, to three dimensions, the delta-function
potential in one dimension.  An excellent explanation of this pseudo-potential
is given by Blatt and Weisskopf \cite{blattweisskopf}.  One of the most
far-reaching applications of the pseudo-potential is to the study of many-body
systems \cite{huangyangETAL}.

Once the Fermi pseudo-potential in three dimensions is well understood, it is
natural to ask whether similar considerations can be applied in one
dimension.  It turns out that, going from three dimensions back to one
dimension, the interaction at a point is not limited to the original
delta-function potential.  In fact, the most general potential for the
interaction at one point $x = 0$ consists of three terms, the delta-function
potential together with two Fermi pseudo-potentials \cite{WuCERN097}:
\begin{equation}
V(x,x') = g_1\delta(x)\delta(x')
+g_2[\delta'_p(x)\delta(x')+\delta(x) \delta'_p(x')]
+g_3\delta'_p(x)\delta'_p(x').\label{eq:9} 
\end{equation} 
Here $V(x,x')$ is defined by
\begin{equation}
(V\psi)(x)=\int dx'\,V(x,x')\psi(x'),\label{eq:10}
\end{equation}
while $\delta'_p(x)$ is similar to the derivative $\delta'(x)$ of the delta
function and is defined by
\begin{equation} 
\delta'_p(x)f(x)=\delta'(x)\tilde f(x)\label{eq:11}
\end{equation}
with
\begin{equation}
\tilde f(x)=\begin{cases}{\displaystyle f(x)- \lim_{x\to
0+}\,f(x),}&\mbox{for}\
x>0,\\ {\displaystyle f(x)-\lim_{x\to 0-}\,f(x),}&\mbox{for}\ x<0.\end{cases}
\label{eq:12}
\end{equation}
Note that, in Eq.\ (\ref{eq:9}), the $g_1$ and $g_3$ pieces are even while the
$g_2$ piece is odd.

(9) One of the simplest examples for the quantum memory may be obtained as
follows.  First, in order to have scattering processes \cite{pikesabatier2002}
and also interference between the incident and scattered waves, it is
essential to have at least one spatial dimension.  Secondly, the number of
states $|j\rangle$ for the memory must be more than 1; otherwise, the
information in the quantum memory consists only of one phase, and is thus no
more than what is stored in a classical memory.  In the language of potential
scattering, this means that there are coupled channels.  It should perhaps be
mentioned that it has been known for several years that two-channel
Schr\"odinger equations have some properties not expected from the one-channel
case \cite{Khuri}.

To get a simplest example, the number of spatial dimensions and the number of
channels are both chosen to be the smallest allowed values, namely, one
spatial dimension and two channels.  Thus the Schr\"odinger equation under
consideration is
\begin{equation}
-\frac{d^2\Psi(x)}{dx^2} +(V\Psi)(x)=k^2\Psi(x)\label{eq:13}
\end{equation}
on the entire real axis, where $\Psi(x)$ has two components and $V$ is a
$2\times 2$ matrix.  From the knowledge of the Fermi pseudo-potential in one
dimension discussed above, the most useful choice of $V$ for the present
purpose is, with $g_1\ne 0$ and $g_3\ne 0$, 
\begin{equation}
V(x,x')=g_1\delta(x)\delta(x')\sigma_1+g_3\delta'_p(x)\delta'_p(x')
\sigma_3,\label{eq:14}
\end{equation}
where the $\sigma$'s are the usual Pauli matrices.

Since there is no $g_2$ term in Eq.\ (\ref{eq:14}), this $V(x,x')$ is even
under space reflection.  Thus, the even wave and the odd wave do not mix.  A
straightforward calculation gives the following simple results for the
$S$-matrix of these even and odd waves:
\begin{equation}
S_+(k)=\exp[i\sigma_1\phi_+(k)]\quad\mbox{and}\quad S_-(k)=\exp[i\sigma_3
\phi_-(k)],\label{eq:15}
\end{equation}
where
\begin{equation}
\phi_+(k)=-2\tan^{-1}\frac{g_1}{2k}\quad\mbox{and}\quad\phi_-(k)=-2\tan^{-1}
\frac{g_3k}{2}.\label{eq:16} 
\end{equation}
Note that $S_+(k)$ is independent of $g_3$, while $S_-(k)$ is independent of
$g_1$.  Since any element of SU(2) can be expressed as a finite product of
these $S_+(k)$ and $S_-(k)$ with suitably chosen $k$'s, there is an infinite
number of the desired $\psi^{\text{in}}$ as given by (\ref{eq:4}), provided that
a finite sequence of $\psi^{\text{in}}$ is considered as a $\psi^{\text{in}}$. 
Thus, writing presents no problem in this example.

Consider reading next.  Let $\Bigl[\begin{matrix} a_1\\[-3pt]
a_2\\\end{matrix}\Bigr]$ be the normalized quantum state of the memory. Then
scattering, for example, by an odd incident wave leads to the wave function
\begin{equation}
\Psi(x)=\left[\begin{matrix} a_1\\ a_2\end{matrix}\right] e^{-ikx} +\left[
\begin{matrix} a_1e^{i\phi_-(k)}\\
a_2e^{-i\phi_-(k)}\end{matrix}\right]e^{ikx}\label{eq:17}
\end{equation} 
when $x>0$.  Since
\begin{equation}
\Psi(x)^{\dag}\Psi(x)=2[1+2\cos 2kx \cos \phi_-(k) -(|a_1|^2-|a_2|^2)\sin
2kx \sin \phi_-(k)],\label{eq:18}
\end{equation}
this scattering process gives, through the interference term, the quantity
\begin{equation}
A_1=|a_1|^2-|a_2|^2.\label{eq:19}
\end{equation}
Similarly, scattering by an even incident wave gives
\begin{equation}
A_2=a_1^*a_2+a_2^*a_1.\label{eq:20}
\end{equation}
These two quantities, $A_1$ and $A_2$, give the values of $a_1$ and $a_2$
except for a common phase.  To obtain this common phase, it is sufficient to
use a further interference with scattering from the standard state.  In this
way, reading and hence resetting can be accomplished.

(10) More than twenty years ago, Benioff \cite{Ben80} led the way in
applying quantum mechanics to computers.  After his pioneering work, there
followed a number of major contributions \cite{majorcontributions}.  Since
then, the field has flourished, as evidenced by hundreds of more recent
papers, a small sample being \cite{smallsample}.  In the majority of the papers
on quantum computing, spatial dependence is not taken into account.  As a
consequence, the usual model for quantum memory consists of a spin system or
its generalization, and the operations on the quantum memory consist of
applying unitary matrices to the quantum state in the memory.  This prevailing
model has led to a number of impressive results.

The present model differs from this previous one mainly in the presence of one
or more spatial dimensions, making it possible to apply scattering in order to
operate on the quantum memory, including the fundamental operations of writing,
reading, and resetting.  This is much more than a change in language, and the
present model may have some advantages, including the following two.  First,
since a quantum memory is necessarily small in size \cite{moore1965},
scattering is the simplest way to change the content of a quantum memory. 
Secondly, from the point of view of physics, spatial dimensions are necessarily
there, and their presence allows more possibilities of analyzing the quantum
memory.

Because of the prominent role played by scattering in the present model, there
are the incident, scattered, and total wave functions.  They do not have any
analog in the previous model. One of the consequences of this difference is
that the Holevo bound \cite{holevo1973} needs to be modified.  Let us recall
that the phase shift \cite{blattweisskopf} of scattering is determined from the
total wave function; indeed, the process of reading is nothing more than an
application of the usual phase-shift analysis.  In the language of quantum
computing, with an admissible $\psi^{\text{in}}$, both $\Psi^{\text{in}}$ and
$\Psi^{\text{out}}$ are unentangled, but the total wave function is entangled.

(11) In addition to the fundamental operations of writing on, reading, and
resetting a quantum memory, we mention one more operation related to
decoherence \cite{decoherence}.  It is unavoidable even in principle for the
$\psi^{\text{in}}$ to deviate from an admissible one, a fact central to many
issues, including the no-cloning theorem.  When this happens, the out state of
the memory is no longer precisely a pure state; this amounts to decoherence. 
As a quantum memory is operated on more and more times, the decoherence
becomes more and more significant.  It is impossible to use scattering to
remove such accumulated decoherence.

It is therefore essential to have quantum memories periodically ``cleaned
up,'' and this cleaning-up operation is outside of the operations of writing
on, reading, and resetting the memory.  In other words, from the point of view
of physics and engineering, additional operations on the quantum memory must
be provided.  Consider for definiteness the simplest example where there are
two quantum states which are degenerate.  Suppose an external macroscopic field
can be applied to the memory to lift this degeneracy; after a time much longer
than the lifetime of the upper state, the memory goes into a pure state.  If
the external macroscopic field is then removed slowly, the memory remains in a
pure state.  Details will be published elsewhere.

\begin{acknowledgments} 

For the most helpful discussions, we are greatly indebted to John M.
Myers.  One of us (TTW) thanks the Theoretical Physics Division of CERN for
their kind hospitality.
 
\end{acknowledgments}


\begin{thebibliography}{1}
\bibitem{footnote}
This definition of being admissible can
be generalized by restricting the $\sum_j a_j^{\text{in}}|j\rangle$ to be certain
subsets.  This useful generalization is not needed in this
paper.
\end{thebibliography}
\end{document}